\documentstyle[svcon,epsf,psfig,twoside]{report} 
\newcommand{\diff}[2]{\frac{{\rm d} #1}{{\rm d} #2}}
\newcommand{\pdiff}[2]{\frac{\partial #1}{\partial #2}}

\newcommand{\dddot}[1]{\stackrel{\bm{\makebox[0.45ex][c]{.}\makebox[0.45ex][c]{.}\makebox[0.45ex][c]{.}}}{#1}}
\newcommand{\rK}[1]{\left( #1 \right)}           
\newcommand{\eK}[1]{\left[ #1 \right]}           
\newcommand{\first}{$1^{\rm st}$}
\newcommand{\second}{$2^{\rm nd}$}
\newcommand{\third}{$3^{\rm rd}$}

\newcommand{\D}{{\rm d}}
\newcommand{\bm}[1]{\mbox{\boldmath$ #1 $}}
\begin{document} 
\pagenumbering{arabic}
\chapter{Dynamics of Vortices in Two-Dimensional Magnets}
\chapterauthors{Franz G. Mertens and A. R. Bishop}
\begin{abstract} 
  Theories, simulations and experiments on vortex dynamics in
  quasi-two-dimensional magnetic materials are reviewed. These
  materials can be modelled by the classical two-dimensional
  anisotropic Heisenberg model with $XY$ (easy-plane) symmetry.  There
  are two types of vortices, characterized by their polarization (a
  second topological charge in addition to the vorticity): Planar
  vortices have Newtonian dynamics (even-order equations of motion)
  and exhibit strong discreteness effects, while non-planar vortices
  have non-Newtonian dynamics (odd-order equations of motion) and
  smooth trajectories. These results are obtained by a collective
  variable theory based on a generalized travelling wave ansatz which 
  allows a dependence of the vortex shape on velocity, acceleration
  etc.. An alternative approach is also reviewed and compared, namely
  the coupling of the vortex motion to certain quasi-local spinwave
  modes.
  
  The influence of thermal fluctuations on single vortices is
  investigated.  Different types of noise and damping are discussed
  and implemented into the microscopic equations which yields
  stochastic equations of motion for the vortices. The stochastic
  forces can be explicitly calculated and a vortex diffusion constant
  is defined. The solutions of the stochastic equations are compared
  with Langevin dynamics simulations. Moreover, noise-induced
  transitions between opposite polarizations of a vortex are
  investigated.
  
  For temperatures above the Kosterlitz-Thouless vortex-antivortex
  unbinding transition, a phenomenological theory, namely the vortex
  gas approach, yields central peaks in the dynamic form factors for
  the spin correlations.  Such peaks are observed both in combined
  Monte Carlo- and Spin Dy\-na\-mics-Simulations and in inelastic
  neutron scattering experiments. However, the assumption of ballistic
  vortex motion appears questionable.
\end{abstract}  

\section{Introduction}
During the past 15 years an increasing interest in two-dimensional
(2D) magnets has developed. This is a result of (i) the investigation
of a wide class of well-characterized quasi-2D magnetic materials
which allow a detailed experimental study of their properties
(inelastic neutron scattering, nuclear magnetic resonance, etc.), and
(ii) the availability of high-speed computers with the capabilities
for simulations on large lattices. Examples of these materials are (1)
layered magnets \cite{Hirakawa89}, like $\rm K_2CuF_4$, $\rm
Rb_2CrCl_4$, $\rm (CH_3NH_3)_2CuCl_4$ and $\rm BaM_2(XO_4)_2$ with
$\rm M = Co, Ni,\ldots$ and $\rm X = As, P,\ldots$; (2) $\rm CoCl_2$
graphite intercalation compounds \cite{Zabel89}, (3) magnetic lipid
layers \cite{Pom88}, like $\rm Mn(C_{18}H_{35}O_2)_2$. For the first
class of these examples the ratio of inter- to intraplane magnetic
coupling constants is typically $10^{-3}$ to $10^{-6}$.  This means
that the behavior is nearly two-dimensional as concerns the magnetic
properties.  For the second class, the above ratio of the coupling
constants can be tuned by choosing the number of intercalated graphite
layers. For the third class, even monolayers can be produced, using
the Langmuir-Blodgett method.

Many of the above materials have an ''easy-plane'' or $XY$-symmetry.
This means that the spins prefer to be oriented in the $XY$-plane
which is defined as the plane in which the magnetic ions are situated.
The simplest model for this symmetry is the 2D classical anisotropic
Heisenberg Hamiltonian (see Ref.~\cite{Steiner86})
\begin{equation}
\label{h=j}
H = - J \sum_{<m, n>} [S_x^m S_x^n + S_y^m S_y^n + 
    (1 - \delta) S_z^m S_z^n]\,.
\end{equation}
$<m, n>$ labels nearest neighbors of a 2D lattice; usually a square
lattice is used.  The subscripts $x, y$ and $z$ stand for the
components of the classical spin vector ${\bf S}$; the spin length $S$
is set to unity by the redefinition $J \to J/S^2$. $J$ is the magnetic
coupling constant, both ferro- and antiferromagnetic materials were
investigated. The anisotropy parameter $\delta$ lies in the range $0 <
\delta \leq 1$; note that $\delta = 1$ corresponds to the $XY$-model,
not to the so-called planar model where the spins are strictly
confined to the $XY$-plane; this confinement is possible only if one
is not interested in the dynamics. Instead of the exchange anisotropy
in the Hamiltonian (\ref{h=j}), one can also use an on-site anisotropy
term $(S_z^m)^2$, which yields similar results.

There are two kinds of excitations: Spin waves which are solutions of
the linearized equations of motion, and vortices which are topological
collective structures. The vortices are responsible for a topological
phase transition \cite{Berez71, Kosterlitz74} at the
Kosterlitz-Thouless transition temperature $T_{\rm KT}$. Below $T_{\rm
  KT}$, vortex-antivortex pairs are thermally excited and destroyed;
above $T_{\rm KT}$ these bound pairs dissociate and the density of
free vortices increases with temperature. It must be noted that an
order-disorder phase transition is not possible according to the
Mermin-Wagner-Theorem \cite{MW}; the reason is that all long-range
order is destroyed by the long-wave linear excitations in all 1D and
2D models with continuous symmetries.

Both theory \cite{Hikami80} and Monte Carlo simulations \cite{Kawa86}
showed that $T_{\rm KT}$ is only very weakly dependent on the
anisotropy $\delta$, except for $\delta$ extremely close to $0$, when
$T_{\rm KT} \rightarrow 0$. For the materials mentioned above,
coupling constants were estimated from fits to spin-wave theory, and
$\delta$ values are in the range $0.01 - 0.6$, where $T_{\rm KT}$ is
still close to its value for $\delta = 1$.

However, the out-of-plane structure of the vortices (i.~e., the
structure of the $S_z$ components) depends crucially on $\delta$,
while the in-plane structure ($S_x$ and $S_y$ components) remains the
same \cite{Gouvea89}. For $\delta > \delta_c (\approx 0.297$ for a
square lattice \cite{Wysin98}), static vortices have null $S_z$
components, they are termed ''in-plane'' or planar vortices in the
literature. For $\delta < \delta_c$ there are ''out-of-plane'' or
''non-planar'' vortices which exhibit a localized structure of the
$S_z$ components around the vortex center. This structure falls off
exponentially with a characteristic length, the vortex core radius
\cite{Gouvea89}
\begin{equation}
\label{rv=1/2}
r_v = \frac{1}{2} \sqrt{\frac{1 - \delta}{\delta}} \, ,
\end{equation}
in units of the lattice constant $a$. The core radius increases with
diminishing $\delta$, allowing a continuous crossover to the isotropic
Heisenberg model ($\delta = 0$) where the topological excitations are
merons and instantons \cite{Belavin75}, rather than vortices.

\begin{figure}[ht]
\begin{center}
   \unitlength1.0cm
   \begin{picture}(10.5,5.0)
      \put(0.0,4.5){\footnotesize (a)}
      \put(0.0,0.0){\epsfxsize=5cm \epsfysize=5cm
                   \epsffile{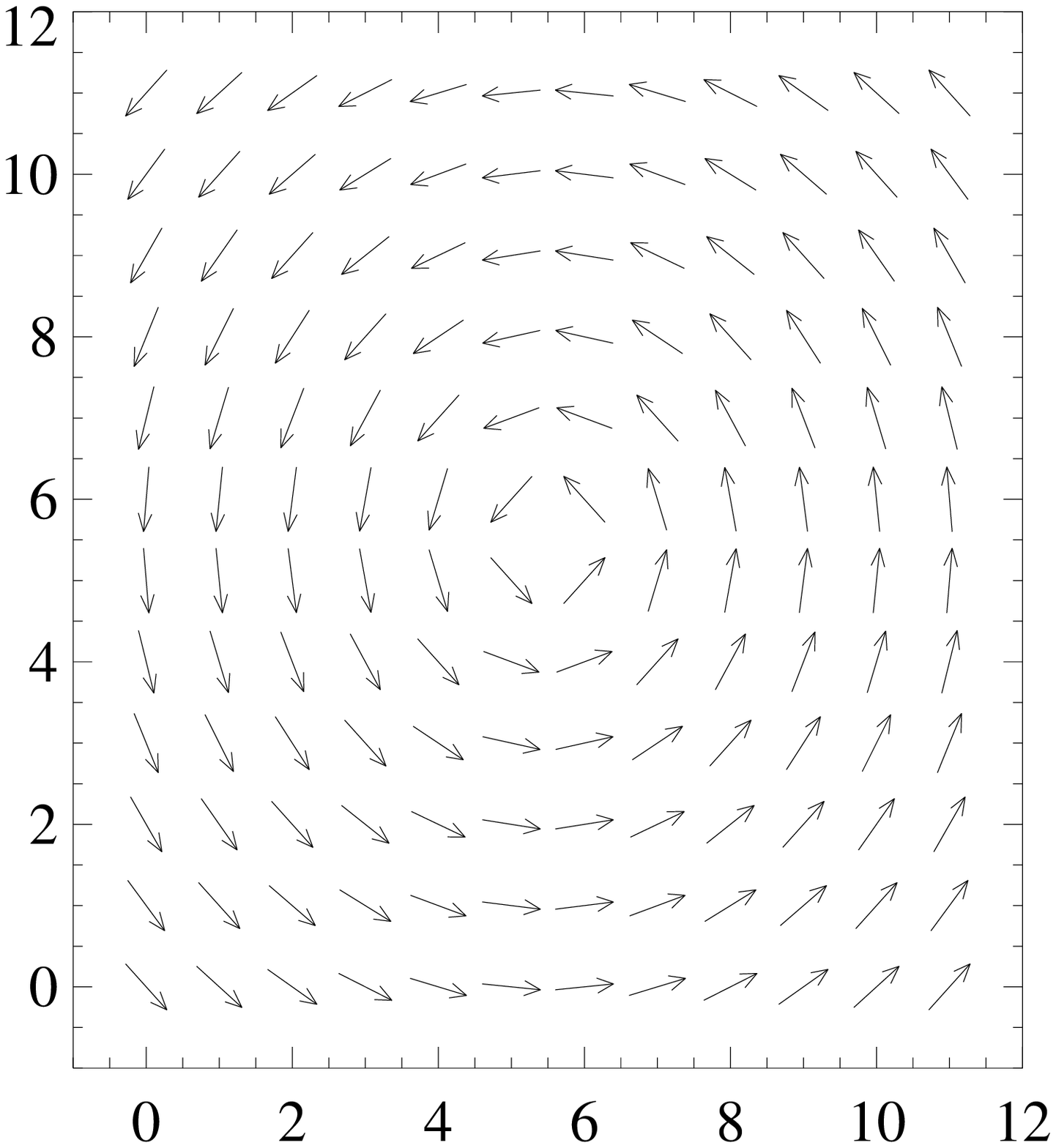}}
      \put(5.5,4.5){\footnotesize (b)}
      \put(5.5,0.0){\epsfxsize=5cm \epsfysize=5cm
                   \epsffile{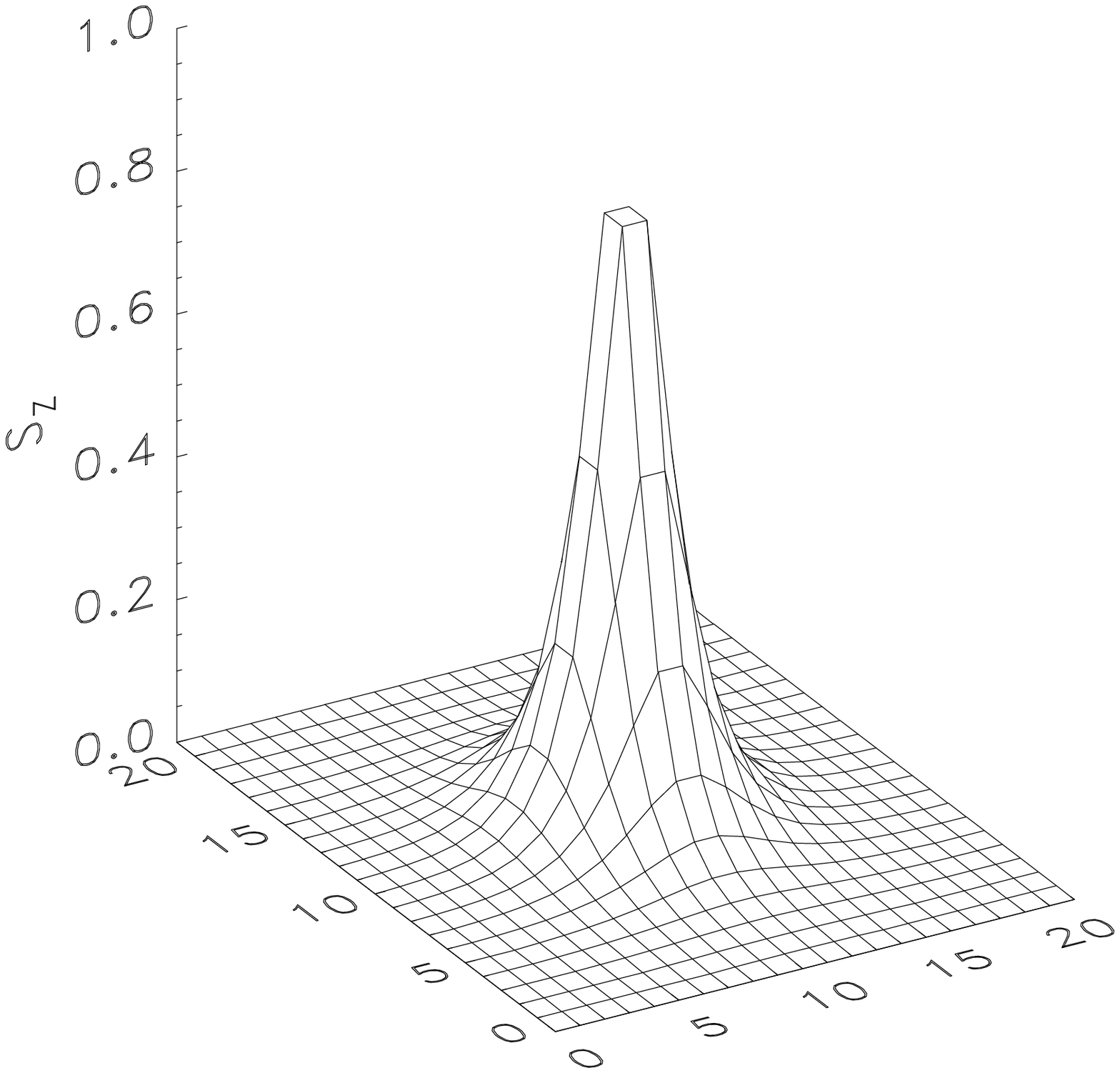}}
   \end{picture}
\end{center}
\caption{(a) In-plane structure of a static planar vortex
  ($q=+1$). (b) Out-of-plane structure of a static non-planar vortex
  with polarization $p=+1$.}
\label{vortexStruct}
\end{figure}

Compared to the vortices in classical fluids \cite{Batchelor67} and
superfluids \cite{Donnely91}, there is an important difference: The
vortices of the easy-plane Heisenberg model carry {\em two}
topological charges, instead of one: (1) The vorticity $q = \pm 1, \pm  2,
\ldots$ which is defined in the usual way: The sum of changes of the
azimuthal angle $\phi = \arctan(S_y/S_x)$ on an arbitrary closed
contour around the vortex center yields $2 \pi q$; if the center is
not inside the contour the sum is zero. In the following only the
cases $q = + 1$ (vortex) and $q = - 1$ (antivortex) will be
considered, the cases $|q| > 1$ would be relevant only for high
temperatures.  (2) The polarization or polarity $p$: For the
non-planar vortices $p = \pm 1$ indicates to which side of the $XY$-plane
the out-of-plane structure points, while $p = 0$ for the planar
vortices (Fig.~\ref{vortexStruct}).

Both topological charges \footnote{From the viewpoint of homotopy
  groups, $q$ and $Q=-\frac{1}{2}qp$ are $\pi_1$- and
  $\pi_2$-topological charges, respectively; see Ref.~\cite{Ivanov95}.}
are crucial for the vortex dynamics, which is governed by two
different forces: (1) 2D Coulomb-type forces ${\bf F}$, which are
proportional to the product of the vorticities of two vortices and
inversely proportional to their distance (assuming that the distance
is larger than $2 r_v$, such that the out-of-plane structures do not
overlap), (2) a ''gyrocoupling'' force \cite{Thiele73,Thiele74}
\begin{equation}
\label{FG=X}
{\bf F}_G = {\bf \dot{X}} \times {\bf G}
\end{equation}
where ${\bf \dot{X}}$ is the vortex velocity, and ${\bf X} (t)$ is the
trajectory of the vortex center.

The force (\ref{FG=X}) is formally equivalent to the Magnus force in
fluid dynamics and to the Lorentz force on an electric charge. The
gyrocoupling vector \cite{Huber82}
\begin{equation}
\label{G=2}
{\bf G} = 2 \pi q p \,{\bf e}_z\,,
\end{equation}
where ${\bf e}_z$ is the unit vector in $z$-direction, does not
represent an external field but is an intrinsic quantity, namely a
kind of self-induced magnetic field which is produced by the localized
$S_z$-structure and carried along by the vortex. For antiferromagnets
${\bf G}$ vanishes.

Since the gyrovector (\ref{G=2}) contains the product of both
topological char\-ges, the dynamics of the
two kinds of vortices is completely different:\\
(1) For planar vortices ${\bf G} = {\bf 0}$, therefore they have a
Newtonian dynamics
\begin{equation}
\label{MX=F}
M{\bf \ddot{X}} = {\bf F}\, ,
\end{equation}
where the vortex mass $M$ will be defined in section 2.2. However, in
the simulations the trajectories are not smooth
due to strong discreteness effects \cite{Volkel91}.\\
(2) Non-planar vortices have smooth trajectories if the diameter $2
r_v$ of the out-of-plane structure is considerably larger than the
lattice constant \cite{Volkel91}; i.~e., if $\delta$ is not close to
$\delta_c$. For steady state motion, the dynamics is described by the
Thiele equation \cite{Thiele73, Thiele74}
\begin{equation}
\label{XxG}
{\bf \dot{X}} \times {\bf G} = {\bf F} \,,
\end{equation}
which was derived from the Landau-Lifshitz equation (section 2) for
the spin vector ${\bf S}^m (t)$. This equation is identical to
Hamilton's equations with Hamiltonian (1.1).

However, for arbitrary motion the Thiele equation (1.5) is only an
approximation because a rigid vortex shape was assumed in the
derivation. If a velocity dependence of the shape is allowed
\cite{Wysin91, Wysin94}, an inertial term $M{\bf \ddot{X}}$ appears on
the l.~h.~s.\ of Eq.~(\ref{XxG}). However, this \second-order equation
could not be confirmed by computer simulations \cite{Volkel94, FGM97}.
The reason will be discussed in section 2; in fact, a collective
variable theory reveals that the dynamics of non-planar vortices can
only be described by odd-order equations of motion. Therefore the
dynamics is {\it non-Newtonian}.

Another interesting topic is the dynamics under the influence of
thermal fluctuations (section 3). Here white noise and damping are
implemented in the Landau-Lifshitz equation. The same type of
collective variable theory as in section 2 then yields stochastic
equations of motion for the vortices. The stochastic forces on the
vortices can be calculated and the solutions of the equations of
motion are compared with Langevin dynamics simulations. In this way,
the diffusive vortex motion can be well understood.

For somewhat higher temperatures (still below $T_{\rm KT}$) the
thermal noise can induce transitions of non-planar vortices from one
polarization to the opposite one. Theoretical estimates of the
transition rate are compared with Langevin dynamics simulations
(section \ref{PolarTrans}).

For the temperature range above $T_{\rm KT}$, so far only a
phenomenological theory exists, the vortex-gas approach
(section~\ref{tempGtKT}). It contains only two free parameters (the
density of free vortices and their r.m.s.  velocity). The theory
predicts ''central peaks'' in the dynamic form factors for both
in-plane and out-of-plane spin correlations. Such peaks are observed
both in computer simulations and in inelastic neutron scattering
experiments, and many of the predicted features are confirmed.
However, a recent {\em direct} observation of the vortex motion in
Monte Carlo simulations has revealed that a basic assumption of the
theory, namely a {\em ballistic} vortex motion, might not be valid.
(In section \ref{tempGtKT} we confine ourselves to the ferromagnetic
case, although easy-plane antiferromagnets were also investigated).

The conclusion in section \ref{Conclusion} will discuss the
ingredients of a theory which can fully explain all observations above
$T_{\rm KT}$.

\section{Collective Variable Theories at Zero Temperature}
\label{zerotemp}
\subsection{Thiele Equation}
\label{Thieleeq}
The spin dynamics is given by the Landau-Lifshitz equation, which is
the classical limit of the Heisenberg equation of motion for the spin
operator ${\bf S}^m$,
\begin{equation}
\label{dS}
\diff{{\bf S}^m}{t} = - {\bf S}^m \times \frac{\partial H}{\partial 
{\bf S}^m} 
\end{equation}
where $H$ is the Hamiltonian, in our case that of the anisotropic
Heisenberg model (\ref{h=j}).  The meaning of Eq.~(\ref{dS}) is that
the spin vector ${\bf S}^m$ precesses in a local magnetic field ${\bf
  B}$, with cartesian components $B_{\alpha} = -\frac{\partial
  H}{\partial S^m_\alpha}$. In spin dynamics simulations (see below),
the Landau-Lifshitz equation is integrated numerically for a large
square lattice, typically with $72 \times 72$ lattice points. As
initial condition a vortex is placed on the lattice and the trajectory
${\bf X}(t)$ of the vortex center is monitored.

This trajectory is then compared with theory, i.~e.\ with the solution
of an equation of motion for the vortex. The standard procedure to
obtain this equation consists in taking the continuum limit ${\bf
  S}^m(t) = {\bf S}({\bf r}, t)$ where ${\bf r}$ is a vector in the
$XY$ plane, and in developing a {\it collective variable theory}.  The
simplest version makes the travelling wave ansatz \cite{Thiele73,
  Thiele74}
\begin{equation}
\label{Srt}
{\bf S}({\bf r}, t) = {\bf S} ({\bf r} - {\bf X} (t))\, ,
\end{equation}
where the functions $S_{\alpha}$ on the r.~h.~s.\ describe the vortex
shape. (Strictly speaking, these functions should bear an index to
distinguish them from the functions on the l.~h.~s.)

As the equation of motion is expected to contain a force, the
following operations are performed with Eq.~(\ref{dS}) which yield
force densities
\begin{equation} \label{5BH}
   {\bf S}\rK{\pdiff{{\bf S}}{X_i}\times \frac{{\rm d}{\bf S}}{{\rm d}t}}=
   -{\bf S}\rK{\pdiff{{\bf S}}{X_i}\times 
      \eK{{\bf S}\times \frac{\delta H}{\delta {\bf S}}}}=
   -S^{2}\frac{\delta H}{\delta {\bf S}}\pdiff{{\bf S}}{X_i}=
      -S^{2}\pdiff{{\cal H}}{X_i} 
\end{equation}
with $i = 1, 2$ and Hamiltonian density $\cal H$.  According to the
ansatz (\ref{Srt}),
\begin{equation}
\label{ds/dt}
\diff{{\bf S}}{t} = \frac{\partial {\bf S}}{\partial X_j} \dot{X}_j
\end{equation}
is inserted, with summation over repeated indices. Integration over
${\bf r}$ then yields the equation of motion
\begin{equation}
\label{GX=F}
{\bf G\dot{X}} = {\bf F}\, ,
\end{equation}
with the force
\begin{equation} \label{8BH}
   F_i = -\int\!\D^2r\, \pdiff{{\cal H}}{X_i} \, ,
\end{equation}
and the gyrocoupling tensor
\begin{equation} \label{eq:Gij}
   G_{ij} =
      \int\!\D^2r\,{\bf S}\pdiff{{\bf S}}{X_i}\times 
                               \pdiff{{\bf S}}{X_j}=
      \int\!\D^2r\,\left\{\pdiff{\phi}{X_i}\pdiff{\psi}{X_j} -
      \pdiff{\phi}{X_j}\pdiff{\psi}{X_i} \right\} \, .
\end{equation}
The expression on the right was obtained by introducing the two
canonical fields
\begin{equation}
\label{phi=}
\phi = \arctan (S_y/S_x), \, \psi = S_z\,
\end{equation}
instead of the three spin components $S_{\alpha}$ with the constraint
$S = 1$. The static vortex structure is \cite{Gouvea89}
\begin{eqnarray}
\label{phi=q}
   \phi_0 &=& q\arctan\frac{x_2'}{x_1'}\\
\label{psi=r_ll_rv}
   \psi_0 &=& p\eK{1-a_1^2\rK{\frac{r'}{r_v}}^2}\quad\mbox{for $r'\ll 
   r_v$} \\
\label{psi=r_gg_rv}
   \psi_0 &=& p a_2\sqrt{\frac{r_v}{r'}} {\rm e}^{-r'/r_v}\quad\mbox{for $r'\gg 
   r_v$} 
\end{eqnarray}
where the constants $a_1$ and $a_2$ can be used for a matching, and 
\begin{equation}
\label{eq325}
   {\bf r}' = {\bf r} - {\bf X} \,.
\end{equation}
The integral then yields
\begin{equation}
\label{Gij2}
G_{ij} = G \epsilon_{ij}\, ,\, G = 2 \pi q p
\end{equation}
where $\epsilon_{ij}$ is the antisymmetric tensor. Interestingly, only
the value $p$ of the $S_z$ component at the vortex {\em center} enters
the final result \cite{Huber82}, i.~e., the out-of-plane vortex
structure needs not to be explicitly known here.

As ${\bf G}$ is antisymmetric, Eq.~(\ref{GX=F}) is identical to the
Thiele equation (\ref{XxG}) with gyrovector ${\bf G} = G{\bf e}_z$.
Since $p = 0$ and thus $G = 0$ for planar vortices, the Thiele
equation is incomplete in this case.  Obviously there must be a
non-vanishing term on the l.~h.~s., this will be an inertial term
(section \ref{vortexmass}).  Moreover, the ansatz (\ref{Srt}), and
thus the Thiele equation, is only valid for steady-state motion when
the vortex shape is rigid (in the moving frame). This includes not
only translational motion with constant velocity $V_0$ but also
rotational motion with constant angular velocity $\omega_0$. Both
types of motion can be obtained by considering {\ two} non-planar
vortices at a distance $2 R_0$ which drive each other by their Coulomb
force $F = 2 \pi q_1 q_2/(2 R_0)$.  For a certain velocity this force
is compensated by the gyrocoupling force (\ref{FG=X}). In fact, since
each vortex carries two charges, there are four physically different
scenarios which represent stationary solutions of the Thiele equation.
They fall into two classes: If the gyrovectors of the two vortices are
parallel (i.  e., $q_1p_1 = q_2p_2$), a vortex-vortex or
vortex-antivortex pair rotates with $\omega_0$ on a circle with radius
$R_0$ around each other, where
\begin{equation}
\label{omR=1/2}
\omega_0R_0^2 = \frac{1}{2}\, \frac{q_2}{p_1} = \frac{1}{2} \, 
\frac{q_1}{p_2}\, .
\end{equation}
If the gyrovectors are antiparallel, the pair performs a parallel
translational motion with velocity $V_0$ and distance $2 R_0$, where
$V_0R_0 = q_2/(2 p_1) = - q_1/(2 p_2)$.

Both kinds of scenarios also appear for vortices in (super)fluids
\cite{Batchelor67, Donnely91}. However, in these contexts there are
only two physically different situations: vortex-vortex rotation and
vortex-antivortex translation.

\subsection{Vortex Mass}
\label{vortexmass}
\begin{figure}[ht]
\centerline{\epsfxsize=8.0truecm \epsffile{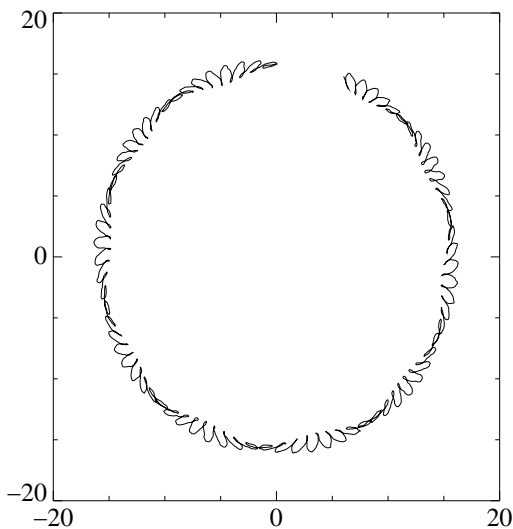}}
\caption{Vortex-antivortex rotation on a circular system with radius
  $L=36$ and free boundary conditions. Only the trajectory of the
  vortex center is plotted, the center of the antivortex is always
  opposite. The mean trajectory is a circle with radius $R_0=15.74$.}
\label{trajectory}
\end{figure}

The above scenarios for magnetic vortices were not tested by computer
simulations until 1994, and the results were very surprising
\cite{FGM94, Volkel94}: Using a square or circular system with free
boundary conditions, two of the four scenarios (vortex-vortex rotation
and vortex-antivortex translation) were confirmed, but not the two
other ones: For vortex-antivortex rotation and vortex-vortex
translation the observed trajectories showed pronounced oscillations
around the trajectories predicted by the Thiele equation (\ref{XxG})
or (\ref{GX=F}), see Fig.~\ref{trajectory}.

In fact, such oscillations had already been predicted \cite{Wysin91,
  Wysin94} by assuming that the vortex shape in the travelling wave
ansatz (\ref{Srt}) depends explicitely on the velocity ${\bf
  \dot{X}}$. This leads to an additional term $\frac{\partial {\bf
    S}}{\partial \dot{X}_j}\, \ddot{X}_j$ in Eq.~(\ref{ds/dt}) and
thus to an inertial term in the Thiele equation
\begin{equation}
\label{mx=gx}
{\bf M\ddot{X}} + {\bf G \dot{X}} = {\bf F}\, ,
\end{equation}
where ${\bf M}$ is the mass tensor
\begin{equation} \label{eq:Mij}
   M_{ij} = 
      \int\!\D^2r\,{\bf S}\pdiff{{\bf S}}{X_i}\times 
                               \pdiff{{\bf S}}{\dot X_j}=
      \int\!\D^2r\,\left\{\pdiff{\phi}{X_i}\pdiff{\psi}{\dot X_j} -
      \pdiff{\phi}{\dot X_j}\pdiff{\psi}{X_i} \right\} \, .
\end{equation}
This integral can be easily evaluated if the vortex is placed at the
center of a circular system with radius $L$ and free boundary
conditions.  The dominant contribution stems from the outer region
$a_c \leq r \leq L$, with $a_c \simeq 3 r_v$, where the velocity
dependent parts of the vortex structure are \cite{Gouvea89, FGM97}
\begin{equation}
\label{psi=q}
\psi_1 = \frac{q}{4 \delta r^2} (x_2\dot{X}_1 - x_1 \dot{X}_2)\, .
\end{equation}
\begin{equation}
\label{eq:phi}
\phi_1 = p (x_1\dot{ X}_1 + x_2\dot{ X}_2)
\end{equation}
Together with the static parts $\phi_0$ from Eq.~(\ref{phi=q}) and
$\psi_0$, which falls off exponentially outside the core
(\ref{rv=1/2}), the rest mass $M$ is obtained
\begin{equation}
\label{eq:M}
   M_{ij} =  M \delta_{ij} \, , \quad
      M  = \frac{\pi q^2}{4\delta}\ln\frac{L}{a_c} + C_{M} \, . \\
\end{equation}
The constant $C_M$ stems from the inner region $0 \leq r \leq a_c$,
where the vortex structure is not well known due to the discreteness,
but $C_M$ is not important for discussion. There is also a velocity
dependent contribution to the mass which is negligible because the
vortex velocities in the simulations are always much smaller than the
spin wave velocity, which is the only characteristic velocity of the
system.

The above vortex mass is consistent with other results in the
literature: Generalizing the momentum of solitons in 1D magnets
\cite{Tjon77}, the vortex momentum
\begin{equation}
\label{P=d}
{\bf P} = - \int\!\D^2 r \, \psi {\bf \nabla} \phi
\end{equation}
is defined and can be shown to be a generator of translations
\cite{Wysin94}.  Then ${\bf P} = M{\bf \dot{X}}$ results, but ${\bf
  P}$ is not a canonical momentum because the Poisson bracket $\{P_1,
P_2\} = G$ does not vanish. For the kinetic energy
$M\dot{{\bf X}}^2/2$ is obtained \cite{Gouvea89}, therefore this
energy and the rest energy $E = \pi \ln (L/a_c) + C_E$ both show the
same logarithmic dependence on the system size $L$. 

For a continuum model of a 2D ferromagnet with uniaxial symmetry and a
magnetostatic field, a slightly different vortex momentum was defined
\cite{Papa91}, namely
\begin{equation}
\label{PPT}
{\bf P}_{PT} = \int\!\D^2r \, {\bf r} \times {\bf g}\, ,
\end{equation}
where ${\bf g} = {\bf \nabla} \phi \times {\bf \nabla} \psi$ represents the
gyrovector density, cf.\ Eq.~(\ref{eq:Gij}) which is related to the
gyrovector as described below Eq.~(\ref{Gij2}). The definition
(\ref{PPT}) depends on the choice of origin of the system and is not a
generator of translations. Nevertheless, if the time derivative of
${\bf P}_{PT}$ is set equal to the force ${\bf F}$ on the vortex, the
generalized Thiele equation (\ref{mx=gx}) is again obtained
\cite{Wysin94}. Therefore the vortex dynamics is qualitatively the
same, as will be discussed now.

Eq.~(\ref{mx=gx}) has the same form as that for an electric charge $e$
in a plane with a perpendicular magnetic field ${\bf B}$ and an
in-plane electric field ${\bf E}$. I.~e., the vortex motion is
completely analogous to the Hall effect: The trajectory is a cycloid
with frequency
\begin{equation}
\label{om=G}
\omega_c = \frac{G}{M}\, ,
\end{equation}
cf.\ the cyclotron frequency $eB/(Mc)$, where $c$ is the speed of
light. It is possible to transform to a frame where the force ${\bf
  F}$ vanishes and the vortex rotates (i.~e., guiding center
coordinates \cite{Jackson75, Papa91, Wysin94}).

The cycloidal trajectories can explain {\em qualitatively} the
oscillations observed in the simulations (Fig.~\ref{trajectory}).
Moreover, the fact that oscillations are observed only for two of the
four scenarios for two vortices (see above) can also be explained: A
generalization of the ansatz (\ref{Srt}) to the case of two vortices
yields two coupled equations of motion with a 2-vortex mass tensor
\cite{Volkelplenum94, Volkel94}. In the case of vortex-vortex rotation
and vortex-antivortex translation there are cancellations in this mass
tensor which prevent oscillations, in agreement with the simulations.

\begin{figure}[ht]
\centerline{\epsfxsize=10cm \epsffile{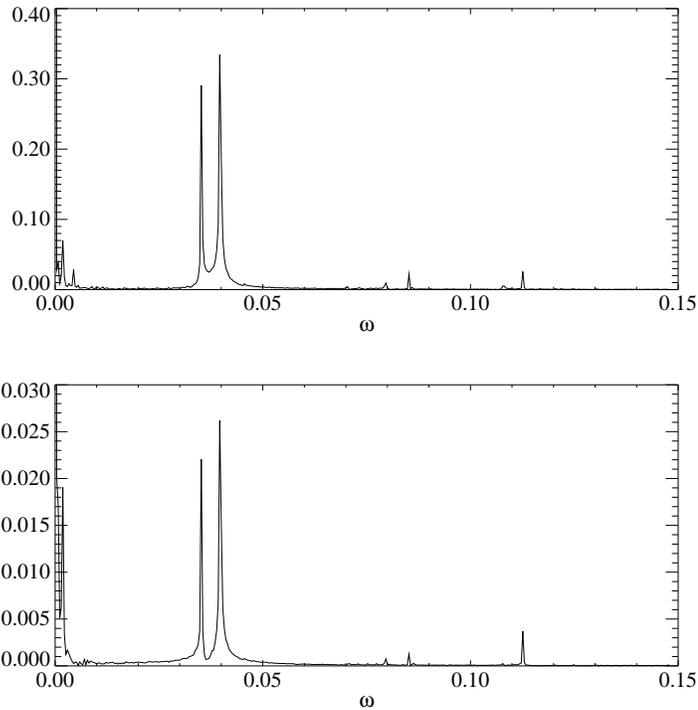}}
\caption{Upper panel: Fourier spectrum of the radial displacements
  $r(t) = R(t) - R_0$ of the vortex in Fig.~\ref{trajectory} from the
  mean trajectory; the integration time is $20000$ in units
  $(JS)^{-1}$.  Lower panel: Spectrum of the azimuthal displacements
  $\varphi(t)=\phi(t)-\omega_0t$ from the mean trajectory, where
  $\omega_0$ is the angular velocity.}
\label{fourierSpec}
\end{figure}

However, a {\em quantitative} comparison of Eq.~(\ref{mx=gx}) and the
simulations reveals two severe discrepancies \cite{Volkel94, FGM97}:\\
(1) The mass $M = G/\omega_c$ which is obtained from Eq.~(\ref{om=G})
by inserting the observed frequency, turns out to be much larger than
predicted by Eq.~(\ref{eq:M}).  Moreover, the dependence on the system
size $L$ is linear, while Eq.~(\ref{eq:M})
predicts a logarithmic dependence.\\
(2) Instead of the {\em one} frequency $\omega_c$ of Eq.~(\ref{om=G}),
{\em two} frequencies $\omega_{1, 2}$ are observed in the spectrum of
the oscillations (Fig.~\ref{fourierSpec}). As $\omega_1$ and
$\omega_2$ are close to each other, a pronounced beat is observed in
the trajectories (Fig.~\ref{trajectory}): The cycloidal frequency is
$(\omega_1 + \omega_2)/2$, but the shape of the cycloids changes
slowly with $(\omega_2 - \omega_1)/2$.

Both discrepancies have nothing to do with 2-vortex effects, because
they also occur in 1-vortex simulations \cite {FGM97}. Here the vortex
is driven by image forces, in analogy to electrostatics. The most
convenient geometry is a circular system with radius $L$ and free
boundaries. In this case there is only one image vortex which has
opposite vorticity, but the same polarization \cite{FGM94, HJS96}.
The radial coordinate of the image is $L^2/R$, where $R$ is the vortex
coordinate.  The vortex trajectory can be fitted very well to a
superposition of two cycloids. This observation will be explained in
the next section.

\subsection{Hierarchy of equations of motion}
\label{hierarchy}
Recently a very general collective variable theory was developed for
nonlinear coherent excitations in classical systems with {\em
  arbitrary} Hamiltonians \cite{FGM97}. In this theory the dynamics of
a single excitation is governed by a hierarchy of equations of motion
for the excitation center ${\bf X} (t)$. The {\em type} of the
excitation determines on which levels the hierarchy can be truncated
consistently: ''Gyrotropic'' excitations are governed by odd-order
equations and thus do not have Newtonian dynamics, e.~g.\ Galileo's
law is not valid. ''Non-gyrotropic'' excitations are so-to-speak the
normal case, because they are described by even-order equations,
i.~e.\ by Newton's equation in the first approximation.

Examples of the latter class are kinks in 1D models like the nonlinear
Klein-Gordon family and the planar vortices of the 2D anisotropic
Heisenberg model (\ref{h=j}). The non-planar vortices of this model
represent the simplest gyrotropic example. 3D models have not been
considered so far.

The basis fo the above collective variable theory is a ''{\em
  generalized travelling wave ansatz}'' for the canonical fields in
the Hamiltonian. For a spin system, as considered in this review, this
ansatz reads
\begin{equation}
\label{S=S}
{\bf S} ({\bf r}, t) = {\bf S} ({\bf r} - {\bf X}, {\bf \dot{X}}, 
{\bf \ddot{X}}, \ldots, {\bf X}^{(n)})\, .
\end{equation}
This generalization of the standard travelling wave ansatz (\ref{Srt})
yields an $(n + 1)^{\rm th}$-order equation of motion for ${\bf X}
(t)$, because Eq.~(\ref{ds/dt}) is replaced by
\begin{equation}
\label{ds=part}
\diff{{\bf S}}{t} = \frac{\partial {\bf S}}{\partial X_j} \dot{X}_j +
 \frac{\partial {\bf S}}{\partial \dot{X_j}} \ddot{X}_j  + \cdots +  
\frac{\partial {\bf S}}{\partial X_j^{(n)}} X_j^{(n + 1)}\, .
\end{equation}
The same procedure as described below Eq.~(\ref{ds/dt}) then yields
the $(n + 1)^{\rm th}$-order equation.

The case $n = 1$ corresponds to the \second-order equation
(\ref{mx=gx}). In one dimension the antisymmetric tensor ${\bf G}$
from Eq.~(\ref{eq:Gij}) vanishes and (\ref{mx=gx}) reduces to a
Newtonian equation. The same is true for the 2D planar
vortices where $G_{ij} = 0$ because $p = 0$ in Eq.~(\ref{Gij2}).\\
The case $n = 2$ yields the \third-order equation
\begin{equation}
\label{AXMX}
{\bf A}\raisebox{0.3mm}{${\bf \dddot{X}}$} + {\bf M} {\bf \ddot{X}} + {\bf G} {\bf
  \dot{X}} = 
{\bf F}({\bf X}) 
\end{equation}
with the \third-order gyrotensor
\begin{equation} 
\label{eq:Aij}
   A_{ij} = \int\!\D^2r\,{\bf S}\pdiff{{\bf S}}{X_i}\times 
      \pdiff{{\bf S}}{\ddot X_j}=
      \int\!\D^2r\,\left\{\pdiff{\phi}{X_i}\pdiff{\psi}{\ddot X_j} -
      \pdiff{\phi}{\ddot X_j}\pdiff{\psi}{X_i} \right\} \, . 
\end{equation}
For the above 1D models and the 2D planar vortices nothing changes
because $A_{ij}$ turns out to be zero (below). But the 2D non-planar
vortices are the first gyrotropic example.

For the calculation of the integral (\ref{eq:Aij}) the acceleration
dependence of the outer region of the vortex is needed \cite{FGM97}
\begin{equation}
\label{psi2}
\psi_2 = \frac{p}{4\delta}\,(x_1 \ddot{X}_1 + x_2 \ddot{X}_2)
\end{equation}
\begin{equation}
\label{phi2=q}
\phi_2 = \frac{q}{8\delta} \,(x_2 \ddot{X}_1 - 
x_1 \ddot{X}_2) \,\ln\frac{r}{{\rm e}L} \,.
\end{equation}
Together with the static parts $\phi_0$ and $\psi_0$ this yields
\begin{equation}
\label{eq:A}  
   A_{ij} =  A \epsilon_{ij} \, , \quad 
      A  = \frac{G}{16\delta}\rK{L^2-a_c^2} + C_{A} \, , \\
\end{equation}
where the constant $C_A$ stems from the inner region.

\begin{figure}[ht]
\centerline{\epsfxsize=10.0truecm \epsffile{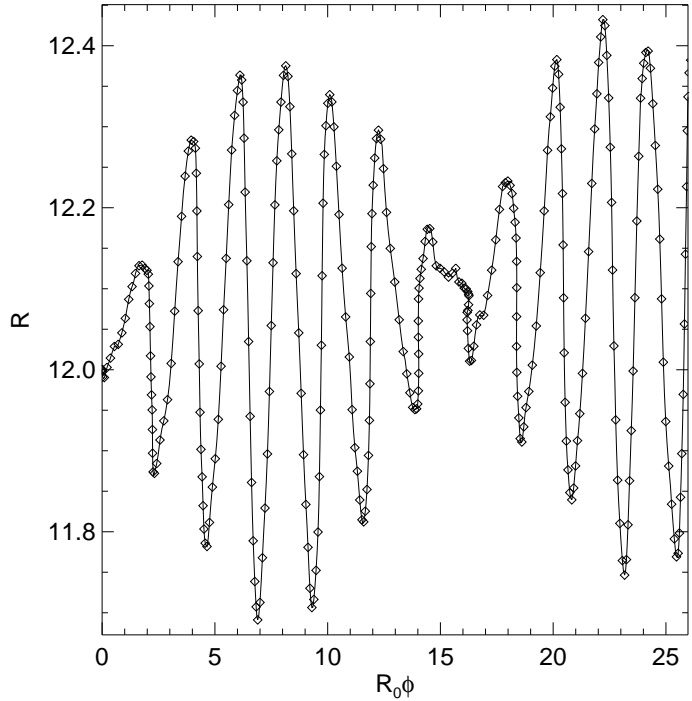}}
\caption{First part of the trajectory of a vortex with $q=p=1$ on a
   circular system with radius $L=36$ and free boundary conditions.
   The small diamonds ($\diamond$) mark the position
   of the vortex in time intervals of $10(JS)^{-1}$.}
\label{trajDetail}
\end{figure}

\begin{figure}[ht]
\centerline{\epsfxsize=8.0truecm \epsffile{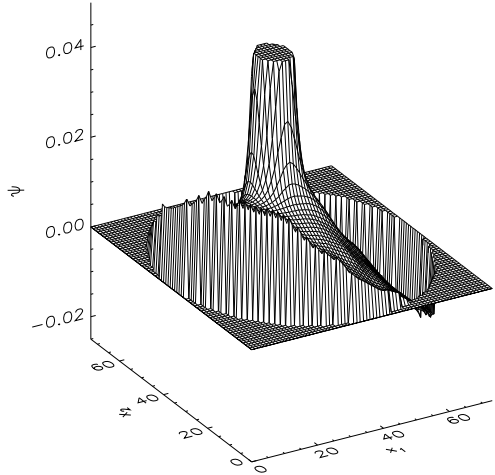}}
\caption{Out-of-plane structure of the vortex at the $7^{\rm th}$
  turning point of the trajectory in Fig.~\protect\ref{trajDetail}.
  Here the acceleration has a maximum and points in the radial
  direction, while the velocity is small and points in the azimuthal
  direction.}
\label{psiStruct1}
\end{figure}

\begin{figure}[ht]
\centerline{\epsfxsize=8.0truecm \epsffile{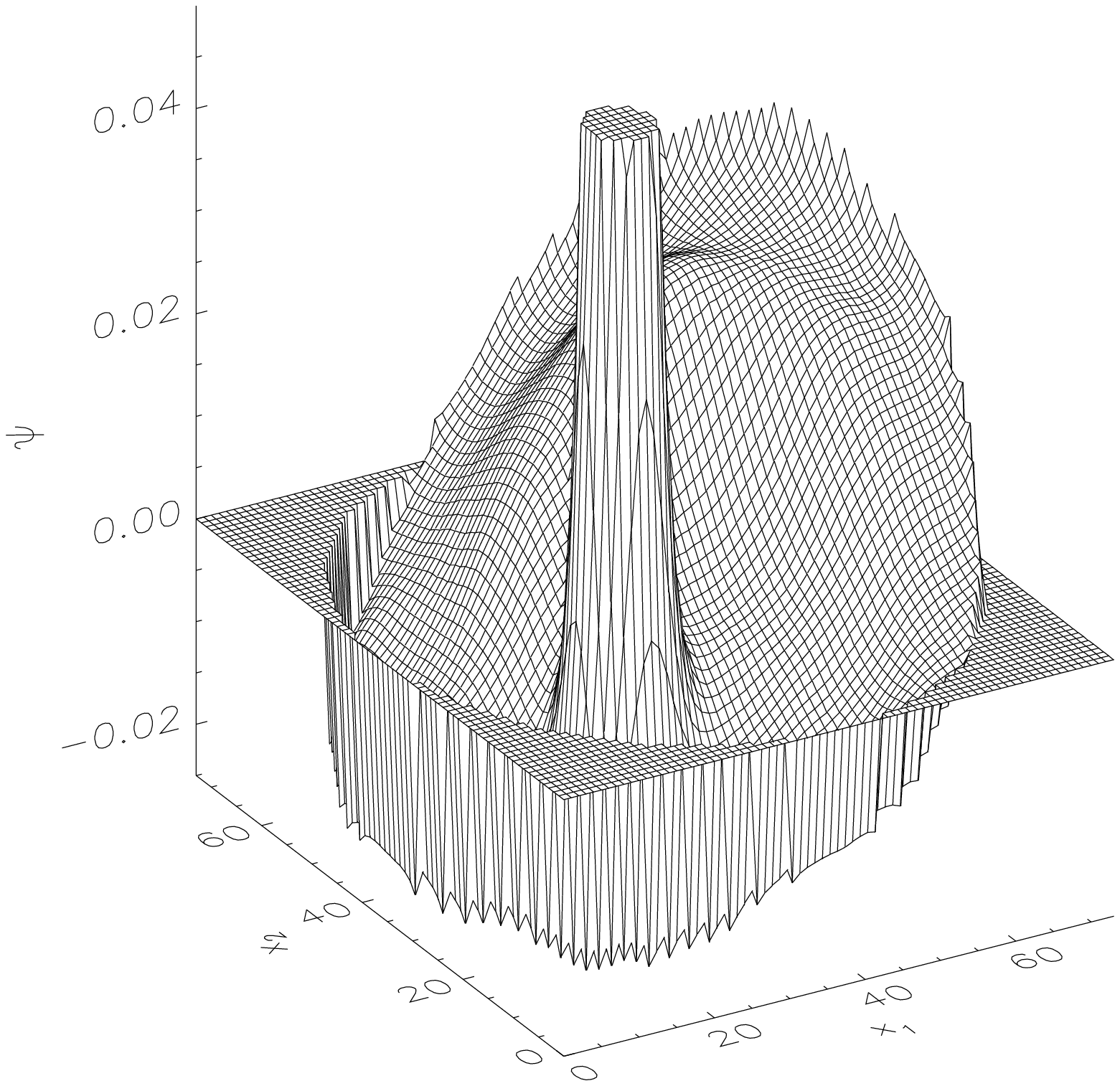}}
\caption{Same as in Fig.~\protect\ref{psiStruct1}, but at the
  $8^{\rm th}$ turning point, where the accelaration points in the
  negative radial direction.}
\label{psiStruct2}
\end{figure}

In the simulations the dynamic in-plane structures $\phi_1$ and
$\phi_2$ cannot be clearly observed because it is difficult to
subtract the static structure $\phi_0$ which varies drastically with
the vortex position.  However, the dynamic out-of-plane structures
$\psi_1$ and $\psi_2$ can be observed and distinguished by looking at
specific points of the trajectory: E.~g., at the turning points in
Fig.~\ref{trajDetail} the acceleration is maximal while the velocity
is small. Figs.~\ref{psiStruct1}, \ref{psiStruct2} confirm the
structure of $\psi_2$ in the outer region, they also show that this
structure oscillates as a whole, with frequency $(\omega_1 +
\omega_2)/2$. An alternative interpretation of this oscillating
structure in terms of certain ''quasilocal'' spinwaves will be given
in section 2.4.

As $\psi_2$ and $\phi_2$ depend linearly on $\ddot{X}_j$ and as
$A_{ij}$ contains derivatives with respect to $\ddot{X}_j$, the
leading contribution to $A_{ij}$ is independent of velocity and
acceleration. There are higher-order terms which are negligible,
though; cf.\ the discussion of the mass. For these reasons the
l.~h.~s.\ of the equation of motion (\ref{AXMX}) is linear.

The radial Coulomb force ${\bf F}$ on the r.~h.~s.\ can be linerarized
by expanding in the small radial displacement $r (t) = R (t) - R_0$
from the mean trajectory which is a circle of radius $R_0$
\begin{equation}
\label{FR}
F(R) = F (R_0) + F'(R_0)r\, .
\end{equation}
The \third-order equation of motion (\ref{AXMX}) then has the
following solutions \cite{FGM97}: A stationary solution and a
superposition of two cycloids
\begin{eqnarray}
\label{ea:r}
r (t) & = & a_1 \cos \omega_1 t + a_2 \cos \omega_2 t\\
R_0 \varphi (t) & = & b_1 \sin \omega_1 t + b_2 \sin \omega_2 t
 \nonumber 
\end{eqnarray}
where $\varphi = \phi - \omega_o t$ is the azimuthal displacement.
The results for $\omega_{1, 2}$ can be considerably simplified for
$R_0 \ll L$, which is the case in the simulations. The frequencies
$\omega_{1, 2}$ form a weakly-split doublet. The mean frequency
depends on $A$, but not on $M$:
\begin{equation}
\label{omsq}
\bar{\omega} = \sqrt{\omega_1 \omega_2} = \sqrt{G/A} \sim 1/L
\end{equation}
for large $L$. Contrary to this, the splitting of the doublet
\begin{equation}
\label{Deltao}
\Delta \omega = \omega_2 - \omega_1 = M/A \sim \frac{\ln L}{L^2}
\end{equation}
is proportional to the mass.

As $G = 2 \pi q p$ is known from Eq.~(\ref{Gij2}), the last two
equations are sufficient to determine $M$ and $A$ by using the
frequencies $\omega_{1,2}$ observed in the simulations. In Ref.
\cite{FGM97} the data for lattice sizes $L = 24 \ldots 72$ were
extrapolated for $R_0 \rightarrow 0$ by using several runs for small
$R_0/L$, because the theoretical predictions (\ref{eq:M}) and
(\ref{eq:A}) were made for $R_0 = 0$. For the anisotropy $\delta =
0.1$ the data for $A$ are well represented by $A = CL^{\alpha} + A_0$
with $\alpha = 2.002,~C = 4.67$ and $A_0 = 40$. This agrees rather
well with the $L^2$-term in (\ref{eq:A}), the constant $C_A$ cannot be
tested. However, $M \approx 15$ is nearly independent of $L$, in
contrast to the logarithmic dependence in Eq.~(\ref{eq:M}). This can
be connected to the observation that the velocity dependent part
$\psi_1$ seems to approach an $L$-independent constant at the boundary,
in contrast to the predicted $1/r$-decay in Eq.~(\ref{psi=q}).

A very interesting point is the discussion of the size dependence of
the different terms in the equation of motion (\ref{AXMX}). As every
time derivative of $X_i$ contributes a factor of $\omega_{1, 2} =
{\cal O} (1/L)$, the orders of the terms are
$A\,\raisebox{0.35mm}{$\dddot{X}_i$} = {\cal O} (1/L), M\ddot{X}_i =
{\cal O} (\ln L/L^2)$ and $G\dot{X}_i = {\cal O} (1/L)$. Therefore the
strong \third-order term cannot be neglected when the weak
\second-order term is retained.  This is the reason for the two severe
discrepancies between the simulations and the predictions of the
\second-order equation (\ref{mx=gx}), which were discussed in section
2.2. However, the neglection of both the \second- and \third-order
terms represents a consistent first approximation, namely, the Thiele
equation (\ref{GX=F}).

The next consistent approximation in the hierarchy is the \third-order
equation (\ref{AXMX}), as discussed above. An investigation of even
higher-order terms in Ref. \cite{FGM97} confirms the conjecture that
only the odd-order equations of the hierarchy represent valid
approximations for gyrotropic excitations (For non-gyrotropic ones all
members of the hierarchy are even).

The odd higher-order equations predict additional frequency doublets
$\omega_{3, 4}$, $\omega_{5, 6}$ etc. These frequencies normally
vanish in the background of the spectrum. However, they become visible
if the simulations are specially designed: For a vortex in the center
of a quadratic system with antiperiodic boundary conditions, all image
forces cancel exactly and only the small pinning forces remain. For
this configuration two additional doublets with strongly decaying
amplitudes were indeed observed \cite{FGM97}.

Finally we mention that cycloidal vortex trajectories have been found
not only in field-theoretic models for magnets \cite{Papa91,Papa95},
but also for (1) neutral and (2) charged superfluids:\\
(1) In the Ginzburg-Landau theory, which describes vortex motion in
thin, neutral, superfluid films, the usual assumption of
incompressibility was dropped \cite{Duan94}. A moving vortex then
exhibits a density profile in the region outside the vortex core. This
profile is similar to the velocity dependent out-of-plane structure
(\ref{psi=q}) of the magnetic vortices, and the consequences are also
similar: There are small-amplitude cycloids \cite{Myklebust96}
superimposed on the trajectories in both 2-vortex scenarios
(vortex-vortex rotation and vortex-antivortex translation, see end of
section \ref{Thieleeq}).\\
(2) For the dynamics of vortices in a charged superfluid the same kind
of superimposed cycloids was found, again for both 2-vortex scenarios
\cite{Stratopoulos96}. Here a field-theoretic model was used, where a
charged scalar field is minimally coupled to an electromagnetic field
and a $\phi^4$-potential is included; this is proposed as a
phenomenological model for a superconductor.

However, compared to the dynamics of magnetic vortices as reviewed in
this article, there is an important difference: For all the models
discussed above, there is only {\em one} cycloidal frequency;
correspondingly the vortex dynamics is governed by \second-order
equations of motion.

It will be interesting to study many other physical contexts in which
topological vortex structures appear (e.~g., dislocations in solids
and flux lattices, vortex filaments in compressable fluids and complex
Ginzburg-Landau models). We expect that the details of the vortex
dynamics differ, depending on the order of the equations of motion,
for instance (see beginning of this section).

\subsection{An alternative approach: Coupling to magnons}
\label{alternative}
The spirit of the {\em generalized travelling wave ansatz} (\ref{S=S})
differs considerably from the well-known {\em standard ansatz}, which
reads for a spin system
\begin{equation}
\label{Srt36}
{\bf S} ({\bf r}, t) = {\bf S}_0 ({\bf r} - {\bf X} (t)) + \bm{\chi}
({\bf {r}}, t)\, .
\end{equation}
Here ${\bf S}_0$ represents the {\em static} structure of a single
nonlinear coherent excitation (a vortex in our case) and $\bm{\chi}$
represents a magnon field (or meson field for other kinds of systems).
Since ${\bf S}_0$ is static, but the shape of the excitation usually
depends on the velocity, a part of the dynamics obviously must be
taken over by the spin waves.

Based on earlier work, this concept has very recently been carried
through \cite{Ivanov98}. First, the magnon modes in the presence of a
single static vortex were obtained by a numerical diagonalization for
relatively small, discrete systems with fixed (Dirichlet) boundary
conditions (BC) \cite{Wysin94a, Wysin95, Ivanov96}.  Analytical
investigations were done for planar vortices in antiferromagnets
\cite{Costa92,Wysin98a} and ferromagnets \cite{Pereira96,Wysin98a}.
These articles demonstrated nontrivial properties of the eigenmodes,
e.~g., the presence of quasi-local (resonance type) \cite{Wysin94a,
  Wysin95} or truly local \cite{Ivanov96} modes.  Moreover, the
relevance of these modes for the vortex dynamics was shown, in
particular the transition between planar and non-planar vortices was
investigated \cite{Wysin94a, Wysin95}.

However, for non-planar vortices all this was based on numerical
diagonalization. But very recently, a general theory of vortex-magnon
coupling was developed for arbitrarily large systems with circular
symmetry and general BC \cite{Ivanov98}. The S-matrix for
vortex-magnon scattering was calculated and expressed by Bessel and
Neumann functions. Using the S-matrix, general formulas for the
eigenfrequencies were obtained, as a function of the parameters and
size of the system, and for different BC, namely for Dirichlet (fixed)
and von Neumann (free) BC.

\begin{figure}[ht]
\centerline{\epsfxsize=11.0truecm \epsffile{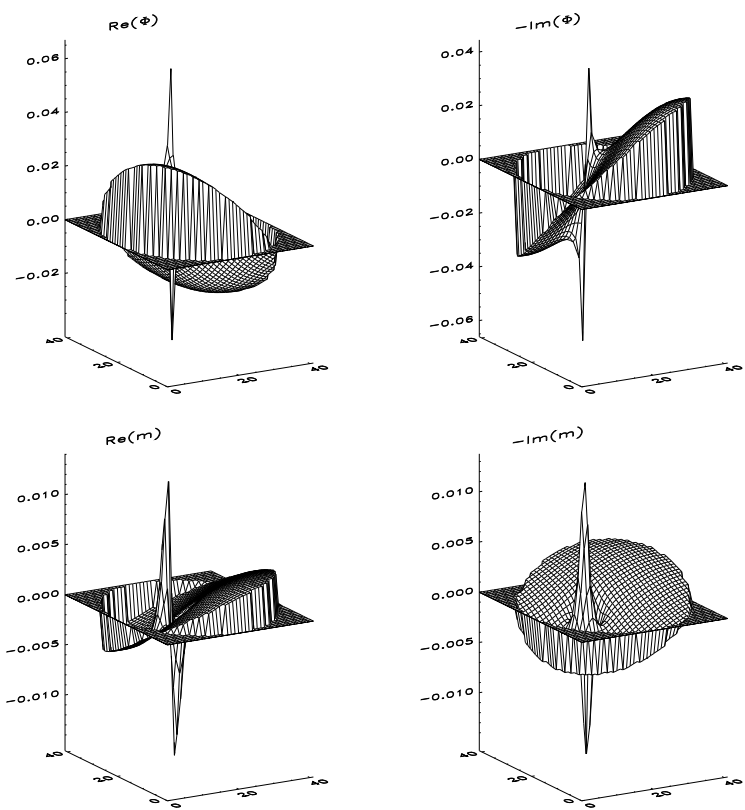}}
\caption{One of the two low-lying quasi-local spin wave modes in a
  complex representation, obtained by a numerical diagonalization for
  a circular system with a vortex in its center, from \cite{HJS96} (In
  this reference $m$ corresponds to our $\psi$).}
\label{spinWave}
\end{figure}

There is a very good agreement between the frequencies of the two
lowest quasi-local modes and the frequencies $\omega_{1, 2}$, which
were observed for the vortex trajectories in spin dynamics simulations
for the discrete system (section (\ref{vortexmass}). The error is only
0.8 \% for the mean frequency $\bar{\omega} = \sqrt{\omega_1
  \omega_2}$ and 4 \% for the splitting $\Delta \omega = \omega_2 -
\omega_1$. This demonstrates very clearly that the {\em cycloidal
  vortex motion} is accompanied by certain magnon modes, namely by
{\em quasi-local modes}. These modes are both extended (like the other
magnon modes) and local, because they exhibit a localized structure
around the vortex center (Fig.~\ref{spinWave}).  The structure of
these modes is qualitatively similar to the dynamic vortex structure,
which oscillates with the rythm of the cycloidal motion
(Figs.~\ref{psiStruct1}, \ref{psiStruct2}).  However, only the
out-of-plane part $\psi$ of this structure can be clearly observed,
and only in the outer region (see section \ref{hierarchy}).

As the frequencies $\omega_{1, 2}$ in the cycloidal vortex motion are
the same as the frequencies of the lowest quasi-local modes, and as
the latter ones were calculated analytically, the parameters $A$ and
$M$ in the \third-order equation of motion (\ref{AXMX}) can also be
calculated analytically \cite{Ivanov98}. The results can be expressed
by the first root $x_1$ of the equation $aJ_1 (x) + b x J_1' (x) r_v/L
= 0$, where $J_1$ is a Bessel function. The values for $a$ and $b$
define the boundary conditions: $a = 0, b = 1$ for free BC and vice
versa for fixed BC.

The general result  
\begin{equation}
\label{A/pi}
A = \frac{\pi}{2 \delta x_1^2} \, L^2
\end{equation}
in the limit of large $L$ specializes to $A = 4.634 L^2$ for $\delta =
0.1$ and free BC, which agrees perfectly with the result $4.67
L^{2.002}$ from the simulations (section 2.3). For fixed BC $A = 1.07
L^2$ is about four times smaller. The general result for the vortex
mass $M$ is much more complicated than Eq.~(\ref{A/pi}) and can be
found in Ref. \cite{Ivanov98}. The numerical value $M = 14.74$ for
$\delta = 0.1$ and free BC agrees well with the value $15$ from the
simulations (section 2.3). However, both results are independent of
the system size $L$, in contrast to the logarithmic $L$-dependence
(\ref{eq:M}) predicted by the integral (\ref{eq:Mij}) in the
collective variable theory, cf.\ the discussion in section 2.3. For
fixed BC, $M = 7.661$ is obtained, about half of the above value for
free BC.

The fact that both the mass $M$ and the factor $A$ in the \third-order
gyrocoupling term depend strongly on the boundary conditions appears
very natural because the vortices are not localized excitations like
solitons in 1D; the in-plane vortex structure falls off with $1/r$.

It is important to note that the higher-lying quasi-local modes could
be calculated by the same methods, and are expected to agree with the
higher-order doublets $\omega_{3, 4}, \omega_{5, 6}, \ldots$ appearing
in the spectrum of the trajectories. Each additional doublet
corresponds to taking into account two additional orders in the
generalized travelling wave ansatz (\ref{S=S}), as discussed in
section \ref{hierarchy}.

Finally, we mention the recent development of a very general
collective variable theory for an arbitrary Hamiltonian $H[\phi,\psi]$
supporting nonlinear coherent excitations, without making any
approximations \cite{HJS96,Schnitzer98}. This theory starts with the
standard ansatz for $\phi({\bf r},t)$ and $\psi({\bf r},t)$, cf.\ 
Eq.~(\ref{Srt36}), but imposes constraints between the meson fields
and the functions $\phi_0({\bf r};X_1,\ldots,X_m)$ and $\psi_0({\bf
  r};X_1,\ldots,X_m)$, which describe the shape of the excitation.
Here the $X_i(t)$ are $m$ collective variables for the position and
the internal modes of the excitation (if there are any).  There are
then $2m$ constraints in order to preserve the correct number of
degrees of freedom. Consequently, the mathematical formalism is based
on the classical limit of Dirac's quantum mechanics for constraint
systems. The equations of motion for the collective variables are a
generalization of Thiele's equation (\ref{GX=F}). The {\em rank} of
the gyrocoupling matrix ${\bf G}$ is used to classify the excitations:
In the case of vanishing ${\bf G}$ the excitations have an effective
mass and exhibit Newtonian dynamics; in the case of regular ${\bf G}$,
the excitations behave like charged, massless particles in an external
magnetic field, similar to the gyrotropic excitations defined in
section \ref{hierarchy}. The above theory is a generalization of
earlier work of Tomboulis \cite{Tomboulis75}, Boesch et al.\ 
\cite{Boesch88}, and Willis et al.\ \cite{Willis90}, which applies
only to ``standard'' Hamiltonians (i.~e., consisting of the sum of
kinetic and potential energy terms). This generalization is necessary,
e.~g., in the case of magnetic systems which cannot be modelled by
standard Hamiltonians.

\section{Effects of thermal noise on vortex dynamics}
\label{Effects}
\subsection{Equilibrium and non-equilibrium situations}
\label{Equi}
As already mentioned in the introduction, below $T_{\rm KT}$
vortex-antivortex pairs appear and vanish spontaneously due to thermal
fluctuations.  But these pairs do not move and therefore give no
contribution to the dynamic correlation functions. Above $T_{\rm KT}$,
some of the pairs unbind and the free vortices can move. In section 4,
this situation will be investigated by a phenomenological theory,
namely the vortex gas approach.

On the other hand, the effect of thermal fluctuations on {\em single}
vortices can be studied by putting a vortex in a thermal bath. This is
a non-equilibrium situation, in fact it takes a long time until
equilibrium is reached: The vortex very slowly approaches the boundary
where it annihilates together with an image antivortex; during this
process spin waves are radiated which are eventually thermalized.

The vortex motion with thermal noise is a random walk process, where a
{\em vortex diffusion constant} can be defined. This offers the
possibility to develop an ab-initio theory for the dynamic form
factors. This can be compared with the phenomenological vortex gas
approach, which assumes, however, a ballistic motion
(section~\ref{tempGtKT}).

\subsection{Collective variable theory and Langevin dynamics simulations}
\label{Collect}
In principle, the generalization of a collective variable theory to
finite temperatures is a straightforward procedure consisting of 4
steps:\\
(1) Thermal noise and damping (because of the fluctuation-dissipation
theorem) are introduced into the microscopic equations.\\
(2) A travelling wave ansatz for a nonlinear coherent excitation is
made which yields equations of motion with stochastic forces acting
on the excitation.\\
(3) The solutions of these stochastic o.~d.~e.'s are used to calculate
the variances of the trajectory, which contain as a factor an
effective diffusion constant for this excitation. The dependences of
this constant on the temperature and other parameters can be
discussed.\\
(4) The predicted variances and the effective diffusion constant are
compared with the same quantities as observed in Langevin dynamics
simulations.

\vspace{0.4cm}
{\it \noindent Step 1: Introduction of thermal noise}
\vspace{0.2cm}

This step is very problematical, although there are many papers in
which the problems are either not appreciated or hidden. There are
basically two major problems:\\
(a) The microscopic equations can often be written in different ways
which are equivalent. In the case of spin systems, the Landau-Lifshitz
equation for the spin ${\bf S}$ is equivalent to the Hamilton
equations for $\phi$ and $\psi$. However, if noise is implemented,
e.~g., by an additive term, the
results may be {\em qualitatively different} (see below).\\
(b) Typically, either {\em additive} or {\em multiplicative} noise can
be used, but many papers do not give a reason why one of the two types
was chosen. However, the results are usually {\em qualitatively
  different} for the two types of noise.

In the case of the vortex dynamics, both major problems have been
investigated in a preprint \cite{Till98}: Additive noise in the
Hamilton equations yields a vortex diffusion constant $D_V$ which
depends {\em logarithmically} on the system size, while in the
multiplicative case $D_V$ is {\em independent} of the system size.
Additive noise in the Landau-Lifshitz equation \cite{Honnef97,
  Tilletal98} yields a diffusion constant with the same logarithmic
term as in the case of additive noise in the Hamilton equations, but
the small constants that appear in addition to the logarithmic term
differ for the two cases because the vortex core gives different
contributions.

Unfortunately, the additive noise in the Landau-Lifshitz equation has
an unphysical feature, namely the spin length is not conserved. In the
collective variable theory this problem is overcome by a constraint,
while in the simulations a renormalization of the spin length is
necessary after every time step\footnote{Technically, this is achieved
  by adding a Lagrange parameter multiplying the constraint
  \cite{Honnef97}. This means that a multiplicative noise term appears
  besides the additive one.}.

However, taking additive noise and using these little tricks is not
really necessary, because there is a better noise term which
eventually leads to the same results, but which is well motivated on
physical grounds and which conserves the spin length \cite{Till98,
  Garanin97}: In the Landau-Lifshitz equation (\ref{dS}) each spin
${\bf S}^m$ precesses in a local magnetic field ${\bf B}$ which is the
gradient of the energy with respect to the spin compononts. This local
field is the only way through which the spin ${\bf S}^m$ can feel any
changes in its environment. Adding a thermal noise term to the local
field thus accounts for the interaction of the spin with magnons,
phonons and any other thermally generated excitations. The stochastic
Landau-Lifshitz equation then reads

\begin{equation}
\label{dS=Sm}
\diff{{\bf S}^m}{t} = - {\bf S}^m \times 
\left[\frac{\partial H}{\partial {\bf S}^m} + {\bf h}^m (t)\right] 
- \epsilon {\bf S}^m \times \diff{{\bf S}^m}{t} \, .
\end{equation}
Because of the cross product with ${\bf S}^m$ this noise is {\it
  multiplicative}. In Ref. \cite{Till98} Gaussian white noise is used:
\begin{eqnarray}
\label{hma}
<h_{\alpha}^m> &=& 0\\
\label{hmat}
<h_{\alpha}^m (t) h_{\beta}^n (t')> &=& D  \delta ^{mn} \delta _{\alpha
  \beta} \delta  (t - t')\, ,
\end{eqnarray}
where $D = 2 \epsilon k_B T$ is the diffusion constant and $\alpha,
\beta$ denote cartesian components. Following Refs. \cite{Thiele73,
  Thiele74, Huber82}, a Gilbert damping with damping parameter
$\epsilon$ was chosen in Eq.~(\ref{dS=Sm}), chiefly because it is
isotropic, in contrast to the Landau-Lifshitz damping \cite{Iida63}.

Strictly speaking, the three equations (\ref{dS=Sm}) do not really
represent Langevin equations, because {\it all} the components of
$d{\bf S}^m/dt$ appear in each equation due to the cross product. To
properly introduce the noise, one first has to group all the time
derivatives on the l.~h.~s.\ of the equation, and only then one can
introduce independent white noise terms for each spin component. But
this procedure eventually produces only a renormalization of the
damping parameter $\epsilon$ in the order of $\epsilon^2$. In the
simulations values of $\epsilon$ in the order of $10^{-3}$ were used,
and therefore the correction factor can even be neglected.

Another, even more important issue is the interpretation of the
sto\-chas\-tic differential equation (\ref{dS=Sm}). As the noise is
{\em multiplicative}, Ito and Stra\-to\-no\-vi\u{c} interpretations do
not coincide, in contrast to the additive noise case
\cite{Kampen80,Gardiner90}. In principle, when thinking of thermal
excitations interacting with the spins, there would be a finite
correlation time which would lead to a colored noise term. Taking
white noise means taking the limit of zero correlation time, and
therefore the stochastic Landau-Lifshitz equation (\ref{dS=Sm}) must
be interpreted in the {\em Stratonovi\u{c} sense}.

As noted in the beginning of this subsection, the remaining three
steps of the stated procedure are straightforward. Therefore we
present here only the results for the case of non-planar vortices.

\vspace{0.4cm}
{\it \noindent Step 2: Stochastic equation of motion}
\vspace{0.2cm}

\begin{figure}[ht]
\centerline{\epsfxsize=7.0truecm \epsffile{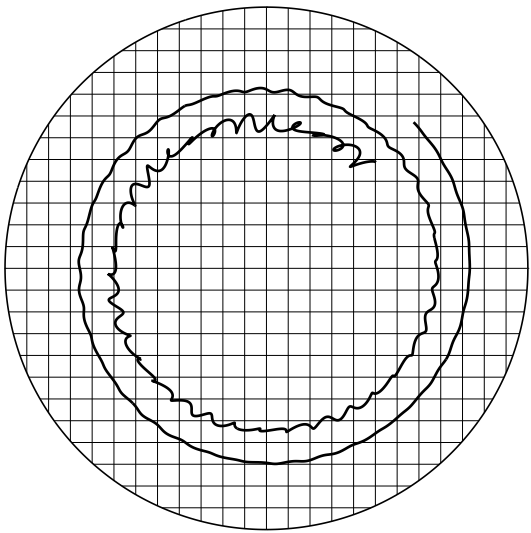}}
\caption{Schematic sketch of the vortex motion as governed by the
  Landau-Lifshitz equation with Gilbert damping. The plot is
  approximate and does not correspond to an actual simulation.}
\label{trajSketch}
\end{figure}

The generalized travelling wave ansatz (\ref{S=S}) up to order $n = 2$
yields the \third-order stochastic o.~d.~e.\
\begin{equation}
\label{AX}
{\bf A} \raisebox{0.3mm}{${\bf \dddot{X}}$} + {\bf M} {\bf \ddot{X}} +
{\bf G} {\bf \dot{X}} = {\bf F}({\bf X}) + {\bf F}^{\rm mult} (t)\,
\end{equation}
The tensors ${\bf A}, {\bf M}$ and ${\bf G}$ are the same as in
Eqs.~(\ref{Gij2}), (\ref{eq:M}) and (\ref{eq:A}), except that all the
vanishing components in these three expressions are now replaced by
damping terms proportional to $\epsilon$. Thus the damping appears at
every order in Eq.~(\ref{AX}). Without ${\bf F}^{\rm mult}$, the
solution is obtained in complete analogy to section \ref{hierarchy}:
Two cycloids are superimposed on an outward spiral on a circle
(Fig.~\ref{trajSketch}).  In the simulations the purpose of the
damping is to dissipate the energy which is supplied to the system by
the kicks of the noise.  Therefore the range of $\epsilon$ (for a
given system size $L$) must be determined in which the cycloidal
frequencies $\omega_{1, 2}$ in the trajectories are not influenced by
the damping \cite{Honnef97}.  The result is the condition $\epsilon L
\ll 6$.

The stochastic force in (\ref{AX})
\begin{equation}
\label{Fist}
F_i^{\rm mult} (t) = \frac{1}{S^2} \int\!\D^2r \frac{\partial {\bf S}}
{\partial X_i} {\bf h} ({\bf r}, t) \,,
\end{equation}
which stems from the multiplicative noise in Eq.~(\ref{dS=Sm}), has
zero mean and the correlation function \cite{Till98}
\begin{equation}
\label{Fist2}
<F_i^{\rm mult} (t) F_j^{\rm mult} (t') > = D \delta_{ij} \delta (t - t') 
\int\!\D^2r \frac{\partial {\bf S}} {\partial X_i} \frac{\partial 
{\bf S}} {\partial X_j} \,.
\end{equation}
Putting the vortex on the center of a circular system of radius $L$
with free BC, the leading contribution to the variance is obtained
\cite{Tilletal98} by using the static vortex structure
(\ref{phi=q}) to (\ref{psi=r_gg_rv})
\begin{equation}
\label{var}
{\rm Var}\left(F_i^{\rm st}\right) = D_V \delta  (t - t')
\end{equation}
with the vortex diffusion constant
\begin{equation}
\label{Dv}
D_V = D \pi \left( \ln \frac{L}{a_c} + C\right)\, ,
\end{equation}
here $C$ stems from the inner vortex region $0 \leq r \leq a_c$ and is
obtained by numerical integration \cite{Tilletal98}.

The variance (\ref{var}) is a remarkable result for two reasons:\\
(a) {\em The stochastic forces acting on the vortex represent additive
  white noise} with an effective diffusion constant. The point is that
this was shown by going from the microscopic level, where {\em
  multiplicative} white noise was implemented, to the level of the
collective variables. This approach is much more satisfying than
assuming {\em ad hoc} a noise term
on the collective variable level.\\
(b) The mean and variance of the stochastic force (\ref{Fist}) turn
out to be the {\em same} as those of the stochastic force
\begin{equation} \label{eq:coll_noise_1}
   F_i^{\rm add} =
   \frac{1}{S^2} \int\!\D^2r \rK{{\bf S} \times \pdiff{{\bf S}}{X_i}}
      \bm{\eta}({\bf r},t)\,,
\end{equation}
which resulted from starting with additive white noise $\bm{\eta}({\bf
  r}, t)$ in the Landau-Lifshitz equation; this noise has already been
discussed in step 1. This equivalence of the two forces means that in
the case of the vortices additive noise and the multiplicative noise
in Eq.~(\ref{dS=Sm}) have the {\em same} effect on the vortex
dynamics. This is a nontrivial result, because normally a qualitative
difference is expected (cf.\ the discussion in step 1). Qualitatively,
this result can be understood by noting that the renormalization of
the spin length in the case of additive noise introduces effectively a
multiplicative noise term, in addition to the additive one (cf.\
footnote in step 1).

\vspace{0.4cm}
{\it \noindent Step 3: Variances of the vortex trajectory}
\vspace{0.2cm}

The inhomogeneous stochastic \third-order o.~d.~e.\ (\ref{AX}) can be
solved \cite{Tilletal98} by a standard Green's function formalism,
because the l.~h.~s.\ is linear and the deterministic force ${\bf
  F}({\bf X})$ on the r.~h.~s.\ can be linearized by expanding in the
displacement ${\bf x}(t) = {\bf X} (t) - {\bf X}^0(t)$ from the mean
trajectory ${\bf X}^0 (t)$ on which the vortex is driven by ${\bf F}$,
cf.\ Eq.~(\ref{FR}). Knowing ${\bf x}(t)$, the variance matrix
\begin{equation}
\label{defvar} 
\sigma_{ij}^2(t)=\langle x_ix_j\rangle - \langle x_i\rangle\langle x_j
\rangle .
\end{equation} 
can be calculated; here $i,j$ denote polar coordinates $R$ and $\phi$.
Each element of the matrix (\ref{defvar}) turns out to consist of 36
terms. In order to facilitate the discussion several simplifications
can be made which give the following results \cite{Tilletal98}:
\begin{equation}
\label{sigma2}
\sigma_{RR}^2 (t) = \frac{D_V}{(2\pi)^2} \left[t + \frac{1}{4 \beta}
  \, (1 - {\rm e}^{-2 \beta t})
- \frac{2}{\bar{\omega}} {\rm e}^{-\beta t} \sin \bar{\omega} t + \frac{1}{4
  \bar{\omega}} {\rm e}^{- 2 \beta t} \sin
  2 \bar{\omega} t\right]
\end{equation}
with $\bar{\omega} = \sqrt{\omega_1 \omega_2}$ and $\beta \simeq
\epsilon/5$ is the damping constant for the vortex motion. For large
times, $t \gg 1/\beta$, only the first term remains and the variance
is linear in time, the standard random walk result. Interestingly,
this result is identical to the one which is obtained by omitting the
\second- and \third-order terms in the stochastic equation of motion
(\ref{AX}). This obviously means that these two terms have
the following effects:\\
(a) They produce the oscillatory parts in Eq.~(\ref{sigma2}), which
are naturally connected
to the cycloidal oscillations in the vortex trajectories,\\
(b) for small times, $t \gg 1/\beta$, the slope of $\sigma_{RR}^2$,
averaged over the oscillations, is larger by a factor of 3/2 compared
to late times.

However, the first effect cannot be observed in the simulations
because the oscillations are hidden in strong discreteness effects
(see next step). For this reason we discuss only the long-time
behavior of the other elements of the variance matrix
\begin{eqnarray}
\label{sigma_R}
\sigma_{R\phi}^2 (t) &=& \frac{1}{2}  \frac{D_V}{(2\pi)^2} \frac{k
    F_0'}{2 \pi} \, t^2 \\
\label{sigma_phi}
\sigma_{\phi \phi}^2 (t) &=&
\frac{D_V}{(2 \pi)^2} \left[t + \frac{1}{3} \,\left( \frac{k
    F_0'}{2 \pi} \right)^2 t^3\right]
\end{eqnarray}
with $k = 1 - F_0/(F_0'R_0)$ and $F_0 = F(R_0)$, $F_0' = F'(R_0)$. The
quadratic and cubic time dependences are {\em non-standard} results
which appear in addition to the standard linear dependence because the
deterministic force field $F(R) = F_0 + F_0'(R - R_0)$ is {\em
  inhomogeneous}. $F$ is a radial force which drives the vortex in the
azimuthal direction, due do the gyrocoupling force (\ref{FG=X}).
Therefore, only the $\phi$-components of $\sigma^2$ are affected:
$\sigma_{R\phi}^2$ acquires a factor $k F_0'/(2 \pi) \cdot t$,
while $\sigma_{\phi \phi}^2$ acquires it twice.

\vspace{0.4cm}
{\it \noindent Step 4: Langevin dynamics simulations}
\vspace{0.2cm}

The stochastic Landau-Lifshitz equation (\ref{dS=Sm}) was solved
numerically for a large circular lattice with free BC and one vortex
driven by a radial image force \cite{Till98,Honnef97,Tilletal98}. The
mean trajectory is an outward spiral. Therefore a small damping
parameter $\epsilon$ was chosen in order to allow long integration
times.

There are several qualitatively distinct temperature regimes. For $0
\leq T < T_3 \approx 0.05$ (in dimensionless units, where $T_{\rm KT}
\approx 0.70$ for the XY-model \cite{Cuccoli95,Evertz96}), two
frequencies are observed in the oscillations around the mean
trajectory which can be identified with the cycloidal frequencies
$\omega_{1, 2}$ in section \ref{hierarchy}. As these frequencies are
constant in the whole regime, the \third-order equation of motion
(\ref{AX}) with {\em temperature independent} parameters can in fact
describe the vortex dynamics.

For $T_3 < T < T_1 \approx 0.3$ the above two frequencies cannot be
observed any longer due to large fluctuations , thus the
\first-order equation of motion is sufficient here. However, in some
of the runs the vortex suddenly changed its direction of motion; this
will be explained in the next subsection.

Finally, for $T > T_1$, a single vortex theory is no longer adequate
because here the probability for the spontaneous appearance of
vortex-antivortex pairs in the neighborhood of the single vortex
becomes too large.

\begin{figure}[ht]
\centerline{\epsfxsize=8.0truecm \epsffile{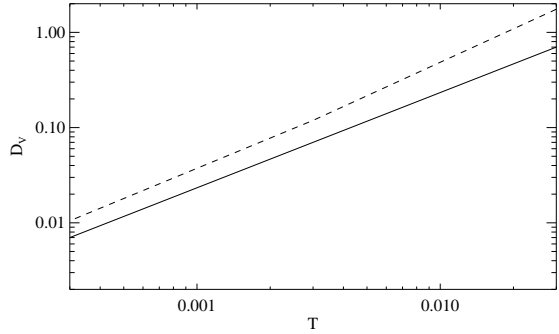}}
\caption{Vortex diffusion constant $D_V$ as a function of temperature,
  for $\epsilon=0.002$ and $L=24$. Solid line: Theoretical results
  from Eq.~(\ref{Dv}); dashed line: Adjusted $D_V$ from fitting the
  theoretical curves for $\sigma^2(t)$ to the simulation data.}
\label{D_effPlot}
\end{figure}
The linear, quadratic, and cubic time dependences in
Eqs.~(\ref{sigma2}) - (\ref{sigma_phi}) are well confirmed by the
simulations. Therefore the factor $D_V$ can be fitted and turns out to
differ from the predicted vortex diffusion constant (\ref{Dv}) only by
a {\em nearly temperature independent} factor of about $1.8$, see
Fig.~\ref{D_effPlot} (The constant $C$ in (\ref{Dv}) was obtained by a
numerical integration over the vortex core \cite{Honnef97}). This
agreement is amazingly good, taking into account that the simulations
were performed for a discrete lattice while the theory works in the
continuum limit and uses additional approximations, like the expansion
(\ref{FR}) of the Coulomb force.

\begin{figure}[ht]
\centerline{\epsfxsize=8.0truecm \epsffile{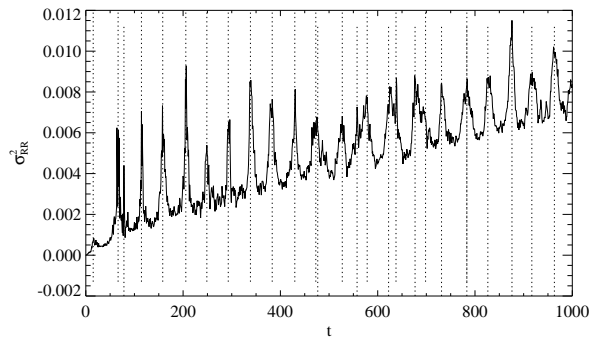}}
\caption{Variance of the radial vortex coordinate averaged over 2000
  realizations, for $T=0.003$, $\epsilon=0.002$, and $L=24$. The
  dashed vertical lines indicate the times at which the vortex center
  moves over ridges of the Peierls-Nabarro periodic potential.}
\label{discrete}
\end{figure}
Although discreteness effects are hardly visible in the trajectories
of non-planar vortices at zero temperature \cite{Volkel91}, see also
Fig.~\ref{trajectory}, these effects are very important at finite
temperatures: Fig.~\ref{discrete} clearly demonstrates that the {\em
  variance} of the trajectory shows a pronounced peak whenever the
vortex center moves over a ridge of the Peierls-Nabarro potential. Due
to this effect, the predicted cycloidal oscillations (\ref{sigma2}) in
the variance cannot be observed.

\subsection{Noise-induced transitions between opposite polarizations}
\label{PolarTrans}
As already mentioned above, a vortex in a thermal bath can suddenly
change its direction of motion on its outward spiral. A closer
inspection shows that this occurs because, opposite to the vorticity
$q$, the polarization $p$ is not a constant of motion for a {\em
  discrete} system: The out-of-plane vortex structure can flip to the
other side of the lattice plane due to the stochastic forces acting on
the core spins. Then the direction of the gyrovector (\ref{G=2}) is
reversed and according to the Thiele equation (\ref{XxG}) the
direction of motion is reversed as well (The same holds for the
\third-order equation (\ref{AXMX}) because ${\bf A} \sim {\bf G}$ in
Eq.~(\ref{eq:A})).

In a preprint \cite{Gaididei98} the cores of both planar and
non-planar vortices are described by a discrete Hamiltonian, similar
to the one which was used for the study of the instability at
$\delta=\delta_c$ \cite{Wysin94a,Wysin98}. Using the stochastic
Landau-Lifshitz equation (\ref{dS=Sm}) the Fokker-Planck equation is
derived. Its stationary solution exhibits two maxima for the two
possible polarizations of the non-planar vortex and a saddle point for
the planar vortex, if the anisotropy parameter $\delta$ lies in a
certain temperature dependent range. The rate $\kappa$ for the
transition from one polarization to the opposite one is calculated in
analogy to Langer's instanton theory \cite{langer69,htb90}, using the
fact that for $\delta \rightarrow \delta_c$ there is a soft mode among
the normal modes which were obtained numerically for a system with one
vortex \cite{Wysin95}. Taking into account only the four innermost
spins of the core, a very simple result is obtained
\begin{equation}
\label{kappa}
\kappa=\frac{1}{2\pi}\sqrt{(1-\delta)^2-
(1-\delta_c)^2}\,{\rm e}^{-\frac{\Delta E}{k_B T}}
\end{equation}
where $\Delta E$ is the energy difference between the planar and the
non-planar vortex. For $\delta=0.1$, the average transition times
$\tau=\kappa^{-1}$ are 100440, 4837, and 1060 for $T=0.1$, $0.15$, and
$0.2$ respectively. Despite of the crude model for the vortex core,
these values agree rather well with the transition times from Langevin
dynamics simulations: $\tau=92516$, $4016$, and $825$ for the above
temperatures, with statistical errors of $22\%$, $6\%$, and $10\%$,
respectively.

\clearpage

\section{Dynamics above the Kosterlitz-Thouless transition}
\label{tempGtKT}
\subsection{Vortex-gas approach}
\label{vortexgas}
This is a phenomenological theory which is based on the following
assumptions: Above $T_{\rm KT}$, the {\em free} vortices form a {\em
  dilute} gas and move either {\em diffusively}
\cite{Huber78,Huber80,Huber82} or {\em ballistically}
\cite{Mertens87}. In the former case, only spin {\em autocorrelation}
functions were calculated which lead to dynamic form factors without
wavevector dependence. Therefore we review only the latter case
\cite{Mertens87}:

The density $n_v$ and the r.~m.~s.\ velocity $\overline{v}$ are the
only free parameters. According to Kosterlitz and Thouless
\cite{Kosterlitz74}
\begin{equation}
\label{nv}
   n_v \approx \frac{1}{(2\xi)^2} \,,
\end{equation}
where $\xi$ is the static spin correlation length which diverges with
an essential singularity for $T \rightarrow T_{\rm KT}$ from above.
Therefore the vortex gas is in fact dilute if $T$ is close enough to
$T_{\rm KT}$.

The vortex density is homogeneous only on the average, locally the
distribution is expected to be {\em random}. Therefore the net force
is zero on the average and the distribution around zero is Gaussian,
which yields a {\em Maxwellian} velocity distribution. This also holds
for the non-planar vortices because the velocity is proportional to
the cross product of the net force and the gyrovector, due to the
Thiele equation (\ref{XxG}).

However, the assumption of a ballistic vortex motion is problematic
(section \ref{MCmotion}). Under this assumption the dynamic spin-spin
correlation functions
\begin{equation}
\label{spinSpinCorr}
   S_{\alpha\alpha}({\bf r},t)=
      \langle S_{\alpha}({\bf r},t)S_{\alpha}(0,0) \rangle \,, 
   \alpha=x,y,z \,,
\end{equation}
and their space-time Fourier transforms, namely the dynamic form
factors $S_{\alpha\alpha}({\bf q},\omega)$, can be calculated
analytically \cite{Mertens87}.

There is an important difference between in-plane correlations
($\alpha=x,y$) and out-of-plane ones ($\alpha=z$): As the in-plane
vortex structure is {\em not} localized (it has no spatial Fourier
transform), $S_{xx} = S_{yy}$ are only {\em globally sensitive} to the
presence of the vortices and the characteristic length scale is their
average distance $2\xi$. The dynamic form factor exhibits a ``{\em
  central peak}'', i.~e.\ a peak around $\omega=0$, with the (squared)
{\em Lorentzian} form
\begin{equation}
\label{Sxx}
   S_{xx}({\bf q},\omega)=
      \frac{1}{2\pi^2}\frac{\gamma^3\xi^2}
          {\omega^2+\gamma^2\eK{1+(\xi q)^2}^2} \,,
\end{equation}
where $\gamma=\sqrt{\pi}\overline{v}/(2\xi)$. Here the core structure
did not enter the result, because the theory was worked out on a
length scale much larger than the core radius $r_v$. The integrated
intensity $I_x(q)$ of (\ref{Sxx}) is {\em inversely} proportional to
the density $n_v$, therefore the motion of the vortices actually {\em
  destroys} correlations.

For the out-of-plane correlations $S_{zz}$ the situation is completely
different: The non-planar vortices have statically the localized $S_z$
structure (\ref{psi=r_ll_rv}), (\ref{psi=r_gg_rv}) which has a spatial
Fourier transform, namely the static form factor $f(q)$.  Therefore
$S_{zz}$ is {\em locally sensitive} to the vortices, i.~e., to their
size and shape. Consequently
\begin{equation}
\label{Szz}
   S_{zz}({\bf q},\omega)=
      \frac{n_v}{4\pi^{5/2}\overline{v}}\frac{|f(q)|^2}{q}
      \exp\eK{-\frac{\omega^2}{(\overline{v}q)^2}}
\end{equation}
contains $f(q)$ and is proportional to the density. This {\em
  Gaussian} central peak simply reflects the Maxwellian velocity
distribution.

For the planar vortices the situation is more complicated because they
do not have a static $S_z$ structure but only a dynamic one, namely
Eqs.~(\ref{psi=q}), (\ref{eq:phi}). Therefore the form factor is
velocity dependent which yields a more complicated result for $S_{zz}$
\cite{Gouvea89}, containing the same Gausssian as in Eq.~(\ref{Szz}).
However, the intensity of this peak is much smaller than that of the
peak (\ref{Szz}), because the dynamic $S_z$ components are about two
orders of magnitude smaller than the static ones \cite{FGM97}.

\subsection{Comparison with simulations and experiments}
\label{compSimExp}
In combined Monte-Carlo (MC) and Spin Dynamics (SD) simulations many
features of the predicted dynamic form factors (\ref{Sxx}) and
(\ref{Szz}) were confirmed:\\
(1) {\em In-plane correlations}: Both for the XY-model ($\delta=1$)
\cite{Mertens87} and for the weakly anisotropic case
($\delta<\delta_c$) \cite{Voelkel92}, the observed $S_{xx}({\bf
  q},\omega)$ exhibits a central peak for $T > T_{\rm KT}$ (but not
for $T < T_{\rm KT}$, as expected). The statistical errors were too
large to decide about the shape. However, the width $\Gamma_x(q)$ and
the integrated intensity $I_x(q)$ of the Lorentzian
(\ref{Sxx}) can be fitted to the data, in this way the free
parameters $\xi$ and $\overline{v}$ are determined. $\xi(T)$ agrees
rather well with the static correlation length \cite{Kosterlitz74}.
$\overline{v}(T)$ increases with $T$ and then saturates; except of a
factor of about 2, this agrees with a result of Huber
\cite{Huber80,Huber82} who assumed a diffusive vortex motion and
calculated the velocity autocorrelation function.

Central peaks were also observed in inelastic neutron scattering
experiments: For $\rm BaCo_2(AsO_4)_2$ \cite{Regnault86} and $\rm
Rb_2CrCl_4$ \cite{Hutchings86} the measurement was performed for only
a few $q$-values. The reported widths of the peaks differ from the
predicted $\Gamma_x(q)$ by factors of about $7$ and $2.5$, resp..
However, in this comparison one has to take into account that the
theory has neglected many features of the real quasi-2D materials:
E.~g., the lattice structure, a pronounced in-plane anisotropy in the
case of $\,\rm BaCo_2(AsO_4)_2$, and quantum effects.

For $\rm CoCl_2$ graphite intercalation compounds the $q$-dependence
of the central peak width was measured \cite{Zabel89} and agrees
qualitatively with $\Gamma_x(q)$ from the peak (\ref{Sxx}).

(2) {\em Out-of-plane correlations}: In MC-SD simulations,
$S_{zz}({\bf q},\omega)$ exhibits a central peak only together with a
spin wave peak which sits on its shoulder (in contrast to $S_{xx}$
where the spin waves are strongly softened for small $q$).

For the XY-model ($\delta=1$), which bears planar vortices, the
intensity of the central peak is expected to be small (see end of
section \ref{vortexgas}). Early simulations with large statistical
errors \cite{Gouvea89} had difficulties to substract the spin wave
peak and reported only upper bounds for the width $\Gamma_z(q)$ and
the intensity $I_z(q)$ of the central peak. Recent
simulations with much better statistics investigated the range
$\delta_c<\delta\le1$. Some papers did not find a central peak 
\cite{Costa96,Evertz96} others did \cite{Gouvea97}.

However, for the weakly anisotropic case ($\delta<\delta_c$) with
non-planar vortices a central peak was indeed found \cite{Costa96}.
Experimentally, $S_{zz}({\bf q},\omega)$ was first not accessible
because of intensity problems. Only the use of spin-polarized neutron
beams made it possible to observe a central peak in $\rm Rb_2CrCl_4$
and to distinguish it from the spin wave peak and from other
contributions \cite{Bramwell88}. The measured width is practically
equal to the width of the Gaussian (\ref{Szz}). However, such a good
agreement seems to be accidental because this material exhibits a
breaking of the rotational symmetry in the XY-plane which is described
by a more complicated Hamiltonian \cite{Hutchings81}. For the same
reason it is not clear whether this material actually belongs to the
case $\delta<\delta_c$.

\subsection{Vortex motion in Monte Carlo simulations}
\label{MCmotion}

The vortex gas approach assumes for single free vortices a diffusive
\cite{Huber78,Huber80} or ballistic \cite{Mertens87} motion. Very
recently this was tested in MC-simulations by monitoring the position
of each vortex in the system (free or bound in a pair) as a function
of time \cite{Dimitrov96,Costa98,Dimitrov98}. The surprising result is
that a single vortex very seldom moves freely over a larger distance.
Normally the vortex travels only one or a few lattice spacings until
it annihilates with the antivortex of a pair which meanwhile appeared
spontaneously in the neighborhood. Another possibility is that the
single vortex docks on a pair or a cluster of pairs and after a while
one of the vortices leaves the cluster. These results confirm
suggestions that vortices cannot move freely for more than a few
lattice spacings, which were made by computing the vortex
density-density correlation function \cite{Costa95,Dimitrov96}.

However, the interpretation of these results is not clear at all. One
possibility is that the vortex gas approach is not valid
\cite{Dimitrov96,Costa98}, but then the striking qualitative agreement
with the central peaks in the simulations and experiments is not
understood. We favour another possibility: Probably only the {\em
  effective} vortex motion is important for the dynamic correlation
function $S_{xx}({\bf r},t)$ (\ref{spinSpinCorr}). This would mean
that it does not matter if a vortex is annihilated with the antivortex
of a pair in the considered time interval $[0,t]$, because the vortex
of that pair continues the travel instead of the original vortex. Only
effective lifetimes are seriously affected. This picture is strongly
supported by the mechanism \cite{Mertens87} which yields the central
peak $S_{xx}({\bf q},\omega)$ in Eq.~(\ref{Sxx}): Looking at the
orientation of one particular spin in the XY-plane, one sees that its
direction is {\em reversed} after the vortex center has gone over this
spin or its neighborhood. Thus $S_x$ and $S_y$ have {\em changed their
  signs} after the vortex is gone, and therefore
\begin{equation}
\label{Sxx=-1^N}
   S_{xx}({\bf r},t) \sim \langle (-1)^{N({\bf r},t)} \rangle
       \,.
\end{equation}
Here $N({\bf r},t)$ is the number of vortices which pass an arbitrary
nonintersecting contour connecting $({\bf 0},0)$ and $({\bf r},t)$.
Obviously, it does not matter if those vortices are replaced by other
vortices during their travel (except maybe if the replacement happens
just when the contour is passed). The evaluation of
Eq.~(\ref{Sxx=-1^N}) leads to the central peak (\ref{Sxx})
\cite{Mertens87}.

\section{Conclusion}
\label{Conclusion}
The vortex dynamics at zero temperature can be well understood by a
collective variable theory: A generalized travelling wave ansatz which
allows for deformations of the vortex shape due to velocity,
acceleration etc., leads to equations of motion. In the case of
non-planar vortices, the trajectories exhibit a superposition of
cycloids which is fully confirmed by spin dynamics simulations.

This collective variable theory can be generalized to finite
temperatures which yields stochastic equations of motion. The vortex
motion is diffusive and agrees well with Langevin Dynamics
simulations. Moreover, the rate of noise-induced transitions between
vortex states with opposite polarization is calculated and agrees with
the simulations. 

However, above the Kosterlitz-Thouless transition temperature there is
so far no theory which can fully explain the central peaks which were
observed both in inelastic neutron scattering experiments and in
combined Monte-Carlo and spin dynamics simulations. A qualitative
agreement is achieved by a phenomenological vortex gas theory, but one
of its assumptions, namely ballistic vortex motion, is questionable.
Probably, both the diffusive character of the vortex motion and
annihilation and creation processes must be incorporated into a theory
which can fully explain the above facts. It is likely that the
situation is similar for all Kosterlitz-Thouless (and many other)
phase transitions.

\vspace{-0.5cm}

\acknowledgments

Many thanks for a very fruitful collaboration go to (in historical
order): Gary Wysin (Kansas State University, USA), Elizabeth Gouv{\^e}a
(Belo Horizonte, Brazil), Armin V\"{o}lkel (Xerox, Palo Alto, USA),
Hans-J\"{u}rgen Schnitzer (Aachen, Germany), Angel S{\'a}nchez, Francisco
Dom{\'\i}nguez-Adame, Esteban Moro (Madrid, Spain), Niels Gr{\o}nbech-Jensen
(Los Alamos, USA), Boris Ivanov, Yuri Gaididei (Kiev, Ukraine), Alex
Kovalev (Kharkov, Ukraine) and Till Kamppeter (Bayreuth, Germany). We
are grateful to Frank G\"{o}hmann (Stony Brook, USA), Dimitre
Dimitrov, Grant Lythe, and Roman Sasik (Los Alamos, USA) for critical
readings of the manuscript and constructive comments.


\end{document}